# Impact of Receiver Antenna Polarization and Resource Scheduler on the Downlink Performance of High Velocity Users in 5G Millimeter Wave Small Cell Technology

Fehima Tajrian
*Islamic University of Technology*
Gazipur-1704, Bangladesh.
Email: Fehimatajrian@iut-dhaka.edu

Md. Asif Ishrak Sarder
*Islamic University of Technology*
Gazipur-1704, Bangladesh.
Email: asifishrak@iut-dhaka.edu

Mariea Sharaf Anzum
*Islamic University of Technology*
Gazipur-1704, Bangladesh.
Email: marieasharaf@iut-dhaka.edu

Moontasir Rafique
*Islamic University of Technology*
Gazipur-1704, Bangladesh.
Email: moontasir@iut-dhaka.edu

Abdullah Bin Shams
*University of Toronto*
Toronto, ON M5S 3G8, Canada.
Email: abdullahbinshams@gmail.com

*Abstract*— With the occupancy of the existing frequency spectrum, the demands of the skyrocketing data traffic paved the path for 5G millimeter Wave (mmWave) technology. The wide frequency spectrum and high directivity can elevate the mean data rate. This performance gain comes at the cost of a higher pathloss which limits the use of mmWave technology to small cells. Also, at such high frequencies, the multipath components and blockages appear as persistent barriers leading to alteration of antenna polarization, affecting resource scheduler performance and throughput degradation. The impact of these barriers for high velocity user equipment's (UEs) in mmWave network is still left to be studied. Using closed loop spatial multiplexing transmission scheme, this study analyses the effect of receiver antenna polarization and types of resource scheduler used in base station on the downlink performance of mobile users (0-120kmph) in the mmWave small cell network. Thorough investigation has been conducted to infer which antenna combination should be advantageous under different scheduling algorithms for high mobility UEs. Our results indicate that the scheduler performance is complimented by the receiver antenna polarization, and by appropriate selection a better downlink performance can be sustained. Also, reception under linear polarization performs better over circular polarization for high velocity UEs.

## I. INTRODUCTION

5G millimeter wave (mmWave) technology is one of the potential contenders for supporting traffic surge of 100-1000 times than 4G[1]. The shorter wavelength of mmWave facilitates it with wide frequency spectrum making it suitable for achieving data rates of several Gbps but not without its challenges [2]. The significant challenge for mmWave is its susceptibility towards propagation loss. Moreover, the limited channel numbers of mmWave make in-channel interference dominant [2] and randomly aligned antennas in a crowded mobile environment result in high polarization mismatch. All these factors contribute to unstable signal to interference plus noise ratio (SINR) inherent to mmWave [1].

The high path loss effects the system link budget. To compensate this loss, highly directional Multiple Input Multiple Output (MIMO) antenna and polarization diversity are required [3]. The polarization of the transmitted signal experiences several reflections and scattering events while reaching the receiver via the multipath channel. These events may alter the original transmitted polarization and significantly reduce the SINR at the receiver. Circular Polarization (CP) and Dual Polarization (DP) are proposed to provide a polarization diversity that improves the SINR and increases the channel capacity [4-6]. Another key problem in mmWave is non-static blockages e.g., pedestrians, moving vehicles etc. This alters the pathloss and affects the performance of the resource scheduler. Adaptive proportional fair scheduler is proposed to circumvent this problem to an extent [7-8]. Radio channel characteristics, propagation model, SINR and data rate performance metrices of mmWave cellular network have been studied in recent researches [9-10]. It is suggested that densification, self-hauling, proper MIMO technique selection can provide better SINR and data rate but the study was limited to static case. Authors in [11-12] demonstrated that by using more receiver antennas over transmission antennas, one can enforce diversity gain to improve the downlink throughput in the 5G network. Given the immense potential of mmWave technology, the performance of the signal polarization, where the reflection and scattering events are more frequent, and resource scheduler for high velocity user-equipment (UE) are yet to be studied.

In this paper, we investigated how the antenna polarization at the receiver and the type of resource scheduler in the base station (BS) effects the downlink performance of a 5G mmWave network for an extensive range of UE velocities (0-120 kmph). We considered linear polarization (LPOL), cross polarization (XPOL), SINR based proportional (PF) and round robin (RR) schedulers. The later schedular periodically. allocates resources to the UE's irrespective of their channel quality.

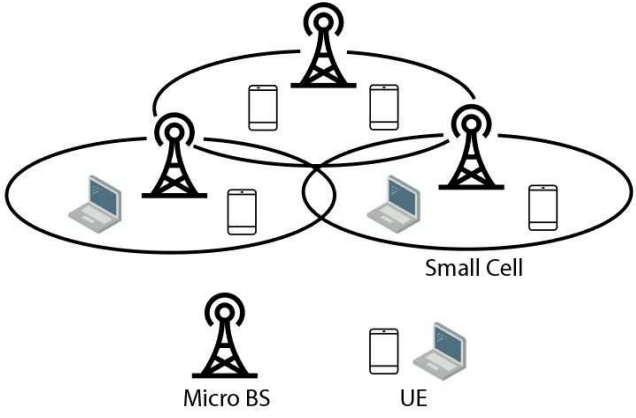

Fig.1: A conventional mmWave small cell network architechture

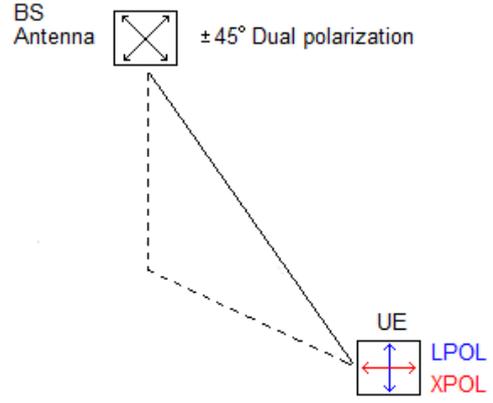

Fig.2: Types of antenna polarization operating in the BS and UE side in the mmWave downlink channel

The rest of the paper is organized as follows. In Section II, various network parameters and network model are presented. Simulation model is discussed in Section III. Results are provided and analyzed in Section IV and finally Section V concludes the whole paper.

## II. System Model

1. *Network Model:*

The simulation was carried out for a model demonstrating a two-tier network consisting of nineteen BS with an inter BS distance ($I_{BS}$) of 500m and every macro BS having tri-sector antenna system. A schematic diagram of the network layout is illustrated by Fig. 1.

2. *Antenna Polarization:*

Linear Polarization (LPOL): The energy radiated by the antenna is focused in a particular intended plane giving it a high directivity with narrow beamwidth.

Cross Polarization (XPOL): Polarization taking place in the perpendicular direction of the intended plane, as shown in Fig.2.

3. *Resource Scheduling:*

This algorithm is present in the MAC layer of the BS and is responsible for the distribution of the available physical resource blocks to the UEs'. Resource allocation is dependant on priority function, P and defined as:

$$P = \frac{T^\alpha}{R^\beta} \quad (1)$$

where T is throughput, R is user data, α and β parameters adjusted according to fairness index [13]. For RR, $\alpha = 0$ and $\beta = 1$. Proportional fair (PF) scheduler allocates resources to users based on user priority to gain optimum throughput at the cost of fairness. Mathematically, the resource allocation of PF algorithm has the following representation [14]:

$$T_k(t+1) = \left(1 - \frac{1}{t_c}\right)T_k(t) + \frac{1}{t_c}\sum_{s=1}^{S} R_{s,k}(t) \quad (2)$$

where $T_k(t)$ is average throughput, $R_{s,k}(t)$ is data rate of user k at sub-band s, $t_c$ is the ), throughput averaging time window that tries to balance between the maximum network throughput and the overall fairness index, and S is the total available sub-bands and commutatively is equal to the operational bandwidth.

4. *Key Performance Indicators:*

*Average UE throughput*: Moving UEs' experiences a rapidly fluctuating SINR which leads to a diverse range of UE throughput [9], therefore we compute the average throughput of all the UEs' to reflect the network performance.

$$T_{avg} = \frac{\sum_{k=1}^{n} T_k}{n} \quad (3)$$

where $T_k$ stands for total throughput for the $k^{th}$ user and n stands for the total number of users.

*Spectral Efficiency*: How accurately data is being transmitted over the given bandwidth (B) is measured from spectral efficiency [13]:

$$S = \frac{\sum_{k=1}^{n} T_k}{B} \quad (4)$$

*Fairness Index (FI)*: The measure of fairness and discrimination in resource allocation between the users is given by FI. According to Jain's FI [15] for *n* users corresponding FI can be expressed as the following equation:

$$J(T) = \frac{[\sum_{k=1}^{n} T_k]^2}{n\,[\sum_{k=1}^{n} T_k^2]} \quad (5)$$

## III. SIMULATION

Using Vienna LTE system level simulator [16], simulation of the 19 base station network was evaluated for UEs randomly scattered in the geometrical area under RR and PF scheduler. At the BS antennas, we used the standard parameters [17]. The overall simulation parameters are in Table I.

TABLE I– SIMULATION PARAMETERS FOR NETWORK MODEL

| Channel model | mmw_Uma_fading |
|---|---|
| Carrier frequency / Bandwidth | 28GHz / 10 MHz |
| MIMO / Transmission mode | 4 × 4 / CLSM |
| BS Height / Receiver height | 25 m / 1.5 m |
| UE per BS | 30 |
| BS antenna azimuth offset | 60° |
| BS transmitter power | 40 Watt |
| BS antenna polarization slant angle | 45° |
| BS antenna electrical down tilt | 90° |
| BS antenna mechanical down tilt | 0° |
| BS antenna mechanical slant | 0° |
| UE antenna polarization | LPOL / XPOL |
| UE antenna polarization slant angle | 0°/ 90° |
| Vertical antenna panel in the BS | 2 |
| Horizontal antenna panel in the BS | 1 |
| Antenna elements per panel | 2 |
| UE velocity | 0-120 kmph |
| Simulation time | 50 TTI |

A dual polarized antenna is implemented at the BS under a 45° slant angle to provide the polarization diversity. In the UE side, receiving antenna is configured with single polarization either LPOL or XPOL, for thorough investigation of the impact of the mentioned transmitting-receiving antenna polarizations on different downlink performance parameters of high velocity users.

## IV.  RESULTS AND DISCUSSION

To observe the distinction in terms of performance, both scenarios corresponding to Linear Polarization (LPOL) and Cross Polarization (XPOL) have been considered separately for the UE reception perspective. As shown in Fig.3, with uniform user velocity increase, a decrement in average UE throughput takes place. With RR scheduler, at low velocity, both LPOL and XPOL show similarity in terms of throughput. However, by maintaining a nearly consistent throughput in all velocities, LPOL outperforms XPOL at high velocities as the latter shows significant decline in throughput when velocity ramps up. Frequent occurrences of scattering, reflection and other multipath events at high velocities may lead to potential phase shift of electric waves giving leeway to unintended polarizations (XPOL) to take place often. Consequently, decreasing the SINR in the system, and therefore hindrance occurs in throughput.

Also, in LTE, generally PF provides higher throughput in comparison to RR, especially at low velocity condition [13]. Contrastingly it can be observed from Fig.3 that in the mmWave channel, because of the extensive lossy condition, the SINR based scheduler (PF) gets hugely affected, whereas RR, with its SINR-independent uniform resource allocation, provides overall better UE throughput for both high and low velocities. While evaluating average UE spectral efficiency, in Fig.4, it can be seen that under PF scheduler, sharp deterioration in performances of LPOL and XPOL occur as the velocity steps up gradually. Under RR, with velocity increase, XPOL performance decreases quite severely in comparison with LPOL. It is clearly visible from Fig.4 that under both PF and RR schedulers, LPOL dominates XPOL at high velocity condition. An acute decline in fairness index is observed in Fig.5 for both XPOL and LPOL under PF scheduler with increase in velocity due to the rapid fluctuations in SINR in mmWave channel especially at high velocity. On the contrary, considering the gradual UE mobility increment, LPOL and XPOL under RR scheduler perform better than the LPOL and XPOL under PF, because of the uniform resource distribution of RR scheduler among UEs without considering channel conditions. In every occasion, LPOL performs significantly better than XPOL under both PF and RR schedulers. So, it can be inferred that under RR scheduler, at low velocity, both LPOL and XPOL can be applicable for UE receiving antenna. In fact, CP can be implemented for the receiving antenna utilizing both LPOL and XPOL configurations as the horizontal and vertical components of CP respectively due to the similarity in their performance at low velocities. For high velocity, the sharp decline in the performance of XPOL renders it unsuitable for implementation. On the contrary, the somewhat consistent performance of LPOL in all velocities deems it fit to be used for the receiving antenna at high velocity. So, the practical solution for RR scheduler should be implementing polarization diversity for UE antenna by providing CP at low velocity and LPOL at high velocity condition. In the case of operation under PF scheduler, because of similarity in throughput delivery of both LPOL and XPOL in every occasion, CP can be implemented for both low and high velocity conditions.

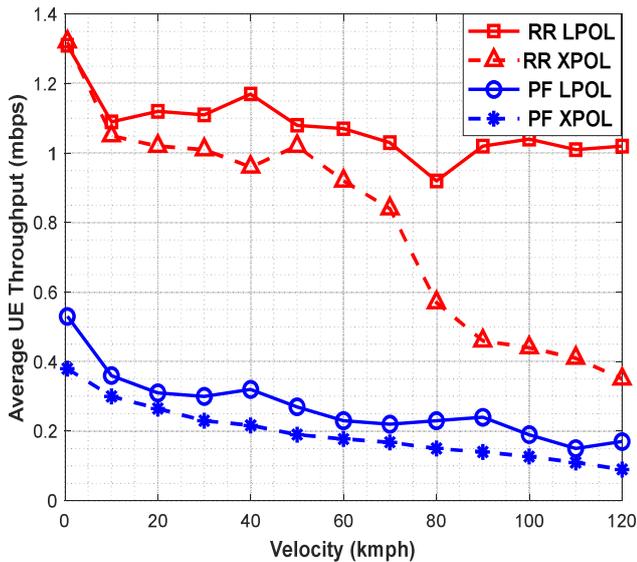

Fig.3: Average UE Throughput vs User Velocity under PF and RR Scheduling algorithms

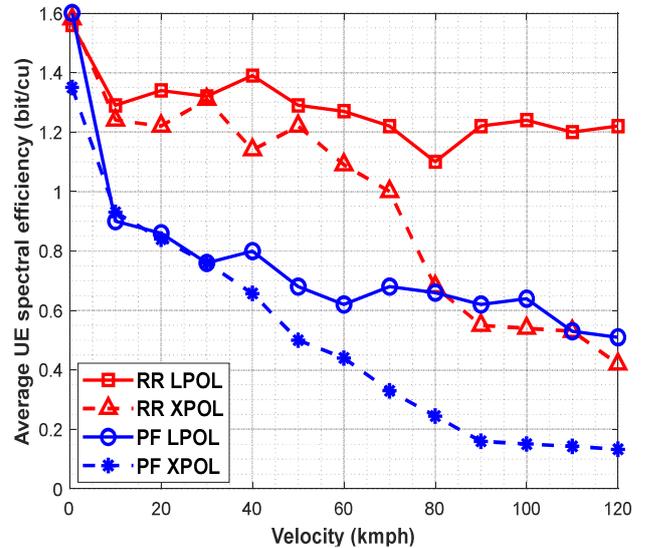

Fig.4: Average UE Spectral Efficiency vs User Velocity under PF and RR Scheduling algorithms

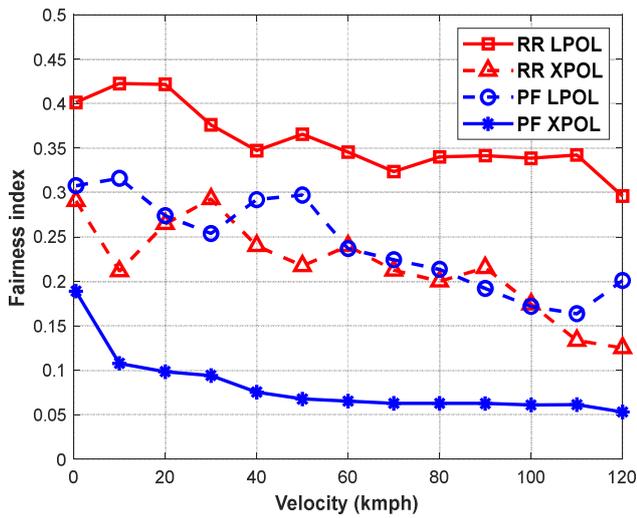

Fig.5: Fairness Index vs User Velocity under PF and RR Scheduling algorithms

## V. CONCLUSION

A large channel bandwidth, followed by a massive data rate, extremely low latency and high directivity, mmWave has become a prominent technology for 5G communication system. Nevertheless, such high operating frequency often cause a significant amount of disruption in the overall network performance. UE mobility increases the amount of scattering and reflection of the signal which further exacerbates the network performance by altering the transmitting signal polarization at the receiver. In this paper, we study the effect of different receiver antenna polarizations operating under various resource scheduling algorithms in the mmWave downlink small cell network under a wide range of UE velocities. Our result demonstrates that channel condition based resource scheduler, PF, is outperformed by channel independent based RR algorithm. The performance difference is more pronounced when the UEs are moving at high velocities. Simulation results also indicate that under the PF scheduler, both LPOL and XPOL provide quite similar kind of throughput irrespective of user velocity but a higher spectral efficiency and fairness was observed for the LPOL. In contrast, under RR scheduler, LPOL and XPOL show similarity in performance at low velocity but as the velocity gradually increases, LPOL delivers far better performance than XPOL in terms of throughput and spectral efficiency. Also, in this case, the fairness index for the LPOL was higher than XPOL. Such observations lead to the conclusion that XPOL signal component suffers significant losses due to scattering and fading for non-static UEs. As the UE moves with a higher velocity the performance of LPOL dominates. Therefore, CP reception can be utilized for low UE velocities, whereas at high velocity, LPOL detection is suitable for UE receiver antenna to ensure extraction of overall better performance offered by mmWave small cell network.